# Second harmonic generation in polycrystalline Zinc Sulfide nanowaveguides.

*Antoine Lemoine1,\*, Lise Morice1, Brieg Le Corre1,2, Antoine Létoublon1, Alex Naïm1, Thomas Batte1, Mathieu Perrin1, Charles Cornet1, Yannick Dumeige1, Christophe Levallois1, Yoan léger1*

[1] Univ Rennes, INSA Rennes, CNRS, Institut FOTON - UMR 6082, F-35000 Rennes, France

[2] Centre de Nanosciences et de Nanotechnologie, CNRS, 91120 Palaiseau, France.

## ABSTRACT

We report the realization of Zinc Sulfide (ZnS) nanowaveguides and the experimental observation of second harmonic generation (SHG) in such structures, demonstrating their potential for integrated nonlinear photonics. ZnS thin films were deposited via RF magnetron sputtering and characterized using atomic force microscopy (AFM), scanning electron microscopy (SEM), X-ray diffraction (XRD), and ellipsometry. The nonlinear optical properties of these films were theoretically analyzed to assess their suitability for second-order nonlinear processes. We detail the fabrication and optical characterization of ZnS nanowaveguides, leading to the experimental observation of SHG in such structures. These findings establish ZnS as a promising platform for nonlinear photonic applications, particularly in compact and integrated frequency conversion devices. This work represents a significant step toward expanding the scope of wide bandgap semiconductors in advanced photonic technologies.

**1. Introduction**

Nonlinear optical phenomena resulting from light-matter interaction have been considerably improved through the development of photonic devices based on a strong confinement of the electromagnetic field which is usually ensured by a guiding structure based on a nonlinear optical material [1]. In the context of octave-spanning spectral conversion, where second-order processes play a crucial role, wide bandgap semiconductors emerge as excellent candidates, particularly for nonlinear conversion into the UV spectral range [2], [3]. This potential can be further optimized by a precise manipulation of the crystal orientation to enhance the nonlinear optical responses. This can be accomplished artificially, like in periodically poled lithium niobate (PPLN) [4] or through the fabrication of orientation-patterned structures to achieve quasi-phase matching [5]. Additionally, it has been demonstrated that naturally occurring polycrystalline structures can be effectively utilized to enhance second harmonic generation (SHG) [6], with potential applications in diverse fields such as neural networks [7] or for realization of optical parametric oscillator [8]. However, the use of this particular crystalline structure is still limited to non-integrated photonic structures. Given these considerations, Zinc Sulfide (ZnS) emerges as a promising candidate for nonlinear photonic applications. ZnS has already found utility in photovoltaic systems [9], photoelectrochemical devices [10],[11] and light sensor in the UV range [12], highlighting its versatility and potential in advanced photonic technologies. However, although cited in pioneering publications on nonlinear optics [13], little research has been carried out ZnS for integrated nonlinear photonics, despite some studies on thin film waveguides for SHG [14], [15] or for integrated electro-optics modulators [16]. Since then, there have been no ZnS-based integrated waveguide designs, notably using dry etching and nanolithography techniques. Yet this material has several advantageous characteristics for nonlinear integrated photonics, such as a high refractive index at telecom wavelengths [17], enabling strong optical confinement, a broad spectral band of transparency from 400 nm to 10 µm [18], high second [19] and third-order nonlinear coefficients [20]. Among the various ZnS deposition techniques, cost-effective methods such as RF magnetron sputtering offer significant technological advantages for practical applications. Moreover, these deposition techniques are cleaner compared to chemical synthesis processes [21], making them highly attractive for technological scalability. A final point of note is the absence of zinc and sulfur from the European Union's list of critical materials, unlike other materials used in nonlinear photonics, such as lithium and niobium, or other III and V elements [22], which ensure a certain degree of sovereignty.

In this work and for the purposes of integrated second order nonlinear optics, we investigate the structural properties of polycrystalline ZnS thin films deposited by RF magnetron sputtering using atomic force microscopy (AFM), scanning electron beam microscopy (SEM) and X-ray diffraction (XRD) measurements. The optical properties of these films are also studied by ellipsometry. We study theoretically the second order nonlinear processes that can be envisaged in such ZnS thin films, in view of their crystalline structure. The fabrication process of the first nanowaveguides based on this new photonic platform is then described, along with their optical characterization. Finally, second harmonic generation and conversion efficiency measurement in these waveguides is reported.

## 2. Fabrication and characterization of sputter deposited ZnS thin films.

In this study, zinc sulfide (ZnS) thin films were deposited by RF magnetron sputtering using 99.9% pure ZnS target from Kurt J. Lesker. To ensure an optimal deposition environment and high quality ZnS layers, a clean vacuum system (residual pressure in the 10-7 mbar range) and stable plasma conditions have been used. A pre-deposition step was performed where the RF plasma power is progressively increased to 85 W in 5 minutes, then a 500 nm thick ZnS layer was deposited onto an oxidized Si substrate over a 30-minute period. For such low plasma power, the heating on the substrate is weak and ZnS can be considered as deposited very close to room-temperature.

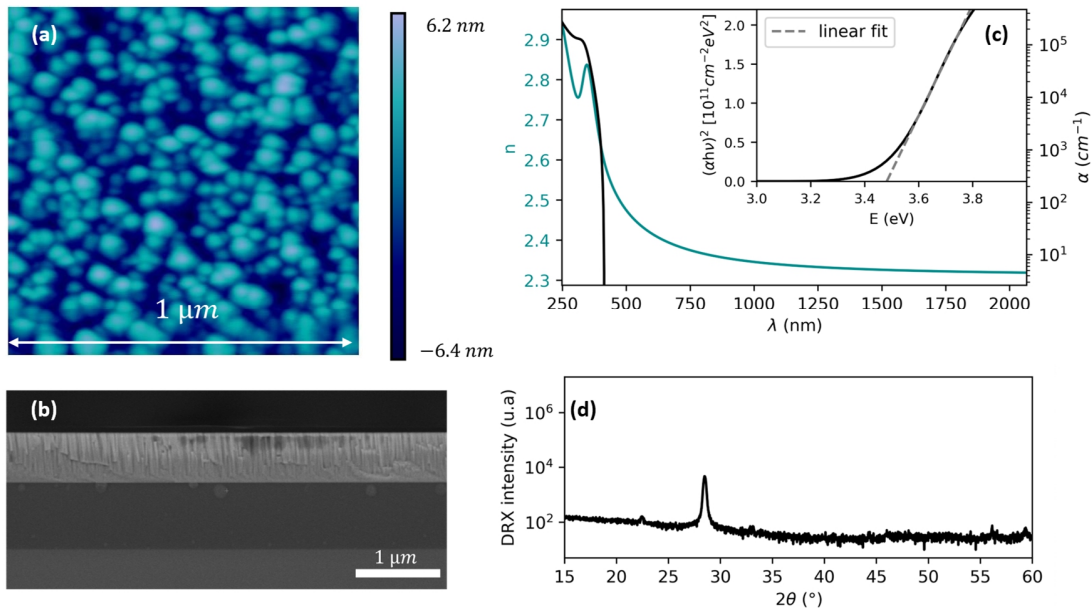

Figure 1 a) 1x1µm AFM image of a ZnS thin film on SiO2 b) SEM image of ZnS thin film on oxidized Si substrate c) Refractive index (n) and absorption coefficient (α) of a ZnS thin film deduced from spectroscopic ellipsometry measurements. Inset shows Tauc plot used for determined bandgap energy d) X-ray diffraction pattern of a ZnS thin film deposited on SiO2/Si.

In these experimental conditions the deposition of polycrystalline ZnS in its cubic (Zinc Blende) phase is well established [23], [24], [25], [26]. This is confirmed by ellipsometry measurements on our samples presented in figure 1.c. The dispersion curves (n, α) of the ZnS layer are deduced from the experimental data using a double Tauc Lorentz model. The insert provides the Tauc plot at the absorption edge, indicating a bandgap energy around 3.5 eV, slightly lower than pure zinc blende ZnS [27] and in very good agreement with other ellipsometry studies carried out on polycrystalline zinc blende ZnS [23], [24]. An absorption shoulder extending up to 450 nm (more visible in the absorption coefficient plot) is also observed. This shoulder suggests the presence of deep-level defects in the ZnS thin films, potentially due to grain boundaries from the polycrystalline structure or residual strain, dependent on film thickness. As presented in Figure 1.a, atomic force microscopy (AFM) analysis of ZnS thin film surfaces revealed a root mean square (RMS) roughness of 2 nm and a coherence length of 35 nm, suitable for the fabrication of photonic structures. The AFM measurements confirm the presence of grains in the layer. Complementary, scanning electron microscopy (SEM) images of the film cross-section (Figure 1.b) confirm a columnar polycrystalline structure. SEM images also show a slight tilt of the columns, which will be of interest in the framework of nonlinear optics in polycrystalline ZnS waveguides. The XRD profiles of Figure 1.d were compared with theoretical patterns from databases such as the Crystallography Open Database. The diffraction peak measured at $2\theta \approx 28.5°$ was found to correspond to zinc blende ZnS oriented along the (111) plane, which is an important parameter for nonlinear processes. By themselves, XRD measurements are not sufficient to confirm the cubic phase, diffraction onto (002) planes of the wurtzite phase sharing the same diffraction angle as cubic (111). Although polymorphism cannot be entirely excluded, the measured bandgap energy shows that the zinc blende structure predominates in this case, in good agreement with the literature [23], [24], [25], [26].

**3. Second harmonic generation in sputter deposited ZnS thin films.**

In a second-order nonlinear medium such as ZnS, the second order nonlinear polarization is defined as:

$$\vec{P}_{NL}^{(2)} = \epsilon_0 \chi^{(2)} : \vec{E}_\omega \vec{E}_\omega$$

Where $\epsilon_0$ corresponds to the permittivity of vacuum, $\chi^{(2)}$ the second order susceptibility tensor and $\vec{E}_\omega$ the incident electric field. Given the Zinc-Blende crystal's orientation of the studied ZnS films, the nonlinear tensor, which is usually expressed in the basis [100],[010],[001] must be transformed to match the [111] growth direction. Firstly, the tensor must be rotated by $\theta_0 = 45°$ around the [001] and then by $\varphi_0 = 54.73°$ around $[1\bar{1}0]$ to finally fit to the $[1\bar{1}0], [11\bar{2}], [111]$ basis, labelled $x', y', z'$ in the following. These two rotations are performed using the corresponding rotation matrices and we label this combined process $\text{Rot}_{\theta_0, \varphi_0}$. In order to simplify the calculation, we choose to work in the non-contracted form and rotate the electric field components using Kronecker products of a rotation matrix of interest with the identity matrix whose dimension agrees with the rank 3 tensor. In the non-contracted form, the second order nonlinear polarization in the [100],[010],[001] basis reads as:

$$\begin{pmatrix} P_x^{(2)} \\ P_y^{(2)} \\ P_z^{(2)} \end{pmatrix} = \epsilon_0 d_{14} \begin{pmatrix} 0 & 0 & 0 & 0 & 0 & 1 & 0 & 1 & 0 \\ 0 & 0 & 1 & 0 & 0 & 0 & 1 & 0 & 0 \\ 0 & 1 & 0 & 1 & 0 & 0 & 0 & 0 & 0 \end{pmatrix} \begin{pmatrix} E_x E_x \\ E_x E_y \\ E_x E_z \\ E_y E_x \\ E_y E_y \\ E_y E_z \\ E_z E_x \\ E_z E_y \\ E_z E_z \end{pmatrix} \quad (1)$$

Since ZnS belongs to crystallographic class $\bar{4}3m$ $d_{14} = d_{25} = d_{36}$ are the only components of $\chi^{(2)}$ which are nonzero [28]. Under these conditions, the second-order nonlinear tensor can be expressed in the basis corresponding to the crystal orientation observed previously, as:

$$\chi^{(2)'} = \text{Rot}_{\theta_0,\varphi_0}^{-1} \times \chi^{(2)} \times \text{Rot}_{\theta_0,\varphi_0} \otimes I \times I \otimes \text{Rot}_{\theta_0,\varphi_0} \quad (2)$$

Due to the preferential orientation of our crystals, we choose to change the basis to match the [111] direction. Firstly, the basis must be rotated by $\varphi_0 = -45°$ around the [001] axis and by $\theta_0 =$ $-54.74°$ around the $[1\bar{1}0]$ axis to finally map to the $[1\bar{1}0], [11\bar{2}], [111]$ basis, labelled $x', y', z'$ in the following. After normalization, the coordinates of the new basis vectors in the old basis are

$$\overbrace{\frac{1}{\sqrt{2}}\begin{pmatrix}1\\-1\\0\end{pmatrix}}^{e_{x'}}; \overbrace{\frac{1}{\sqrt{6}}\begin{pmatrix}1\\1\\-2\end{pmatrix}}^{e_{y'}}; \overbrace{\frac{1}{\sqrt{3}}\begin{pmatrix}1\\1\\1\end{pmatrix}}^{e_{z'}}, \text{ giving } R = \begin{pmatrix} \frac{1}{\sqrt{2}} & \frac{1}{\sqrt{6}} & \frac{1}{\sqrt{3}} \\ -\frac{1}{\sqrt{2}} & \frac{1}{\sqrt{6}} & \frac{1}{\sqrt{3}} \\ 0 & -\frac{2}{\sqrt{6}} & \frac{1}{\sqrt{3}} \end{pmatrix}$$

as the final change-of-basis matrix. This matrix can alternatively be obtained by $R = R_{z,\varphi_0} \cdot R_{x,\theta_0}$, where $R_{z,\varphi_0}$ and $R_{x,\theta_0}$ are traditional rotation matrices. The second-order nonlinear tensor can now be expressed in the new basis adapted to the observed crystal orientation, as:

$$\chi^{(2)'} = R^{-1} \cdot \chi^{(2)} \cdot R \otimes R, \tag{3}$$

where $\cdot$ is the usual matrix product and $\otimes$ the Kronecker product. The contracted second order nonlinear polarization in the $x', y', z'$ basis ($[1\bar{1}0], [11\bar{2}], [111]$) is then given by:

$$\begin{pmatrix}P_{x'}^{(2)}\\P_{y'}^{(2)}\\P_{z'}^{(2)}\end{pmatrix} = \epsilon_0 d_{14} \overbrace{\begin{pmatrix} 0 & 0 & 0 & 0 & -0.57 & 0.81 \\ 0.81 & -0.81 & 0 & -0.57 & 0 & 0 \\ -0.57 & -0.57 & 1.15 & 0 & 0 & 0 \end{pmatrix}}^{\chi^{(2)'}} \begin{pmatrix}E_{x'}^2(\omega)\\E_{y'}^2(\omega)\\E_{z'}^2(\omega)\\2E_{y'}(\omega)E_{z'}(\omega)\\2E_{x'}(\omega)E_{z'}(\omega)\\2E_{x'}(\omega)E_{y'}(\omega)\end{pmatrix} \tag{4}$$

However, due to the polycrystalline nature of ZnS thin films deposited on silica, it is necessary to consider potential disorientation around the preferred [111] growth axis. It is known that for similar polycrystalline materials such as ZnO deposited on silica by similar methods, crystallites can rotate around their growth axis (twist) or be tilted with respect to the same axis (tilt) [29],[30]. It is hypothesized that rotations around the crystallite growth axis can occur over an angle ranging from 0 to $2\pi$. To investigate the influence of the twist on the nonlinear response, we study the evolution of the nonlinear coefficients of the susceptibility tensor for different polarization configurations, with a rotation of angle $\theta$ of the crystal around the [111] axis associated to a rotated basis $x'', y'', z''$ (Figure 2). In this configuration, we obtain several SHG processes for which twist has no effect on the nonlinear response of the crystal such as for example $E_{y''}E_{y''} \rightarrow P_{z''}$, $E_{y''}E_{z''} \rightarrow P_{y''}$ and $E_{z''}E_{z''} \rightarrow P_{z''}$ with a nonlinear coefficient which remains constant as a function of the twist angle $\theta$, meaning that propagation's direction has no influence on the nonlinear response.

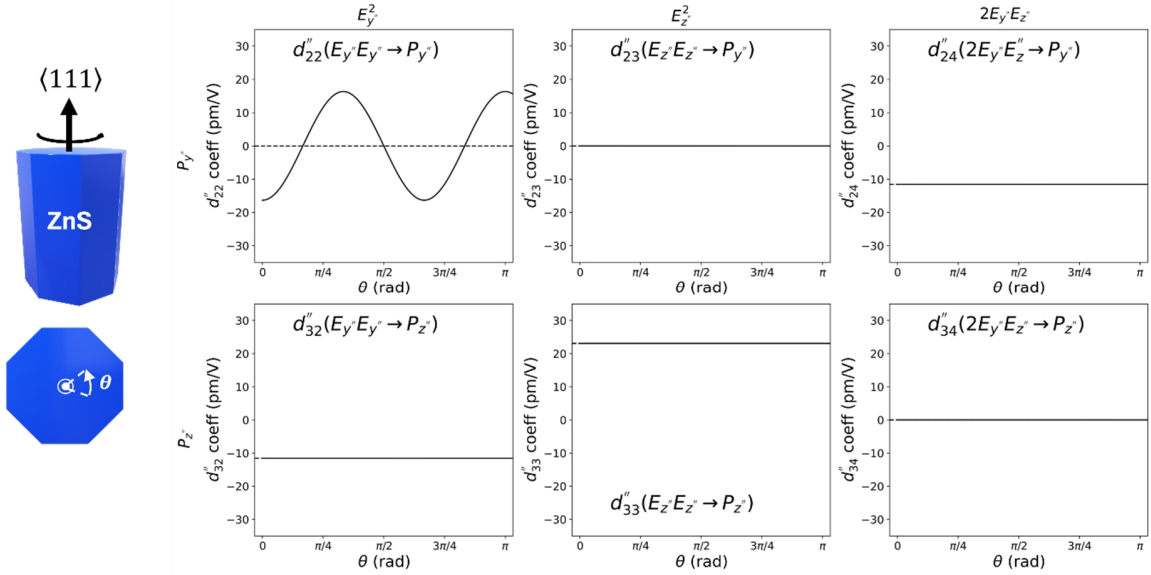

Figure 2 Schematic view of a ZnS column rotating around its growth axis. Evolution of nonlinear coefficients for crystallites twist from 0 to $2\pi$ around [111] crystal axis

This last configuration also features the best nonlinear response due to the perfect orientation of the ZnS dipoles along the incident polarization plane [31]. It will be the configuration explored in the second harmonic generation experiments reported further on. Another potentially advantageous configuration occurs when $d''_{22}$ undergoes a sign inversion of the nonlinear coefficient with each $\pi/3$ crystallite rotation, due to dipole symmetry around the rotation axis (see supporting information for more details), potentially facilitating random quasi-phase matching [32]. In view of the SEM observations presented above, the crystallites appear to be only slightly tilted, but it is not possible to know whether they are reversed by 180° around the $[1\bar{1}0]$ axis from one crystallite to the next. Figure 3 shows that for the practical case with low tilt angle $\varphi$ (blue region), influence is very low on the nonlinear coefficient associated to the polarization configuration of interest. For the SHG processes which are possible when no tilt or twist occur, such as $E_{y''}E_{y''} \to P_{y''}$, $E_{y''}E_{y''} \to P_{z''}$, $E_{z''}E_{z''} \to P_{z''}$ and $E_{y''}E_{z''} \to P_{y''}$, a tilt of 5° results in an average decrease of approximately 1.5% for these four processes. Additionally, Figure 3 shows that in low tilt configurations, processes that were forbidden without tilt, such as $E_{z''}E_{z''} \to P_{y''}$ and $E_{y''}E_{z''} \to P_{z''}$, become possible with a 5° tilt, due to a nonlinear coefficient that changes from 0 to -4.14 pm/V.

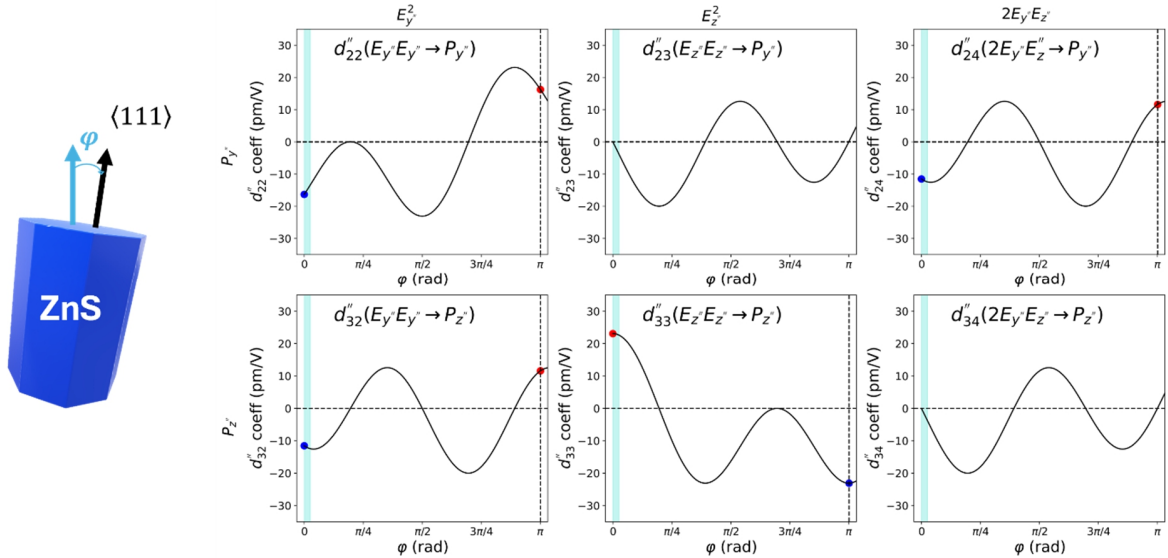

Figure 3 Evolution of nonlinear coefficients in the new basis for φ rotation around [1$\bar{1}$0] for different SHG processes. The values in the blue zone are for low tilt configuration (less than 5°). Blue and red dots are for values with opposite signs.

When the crystalline polarity is reversed ($\varphi = \pi$), the sign of the nonlinear coefficient is reversed for all SHG processes except those for which the nonlinear coefficient is zero in the absence of tilt or twist, for which it returns to zero. This sign inversion, due to dipole reversal, leads to a phase opposition between the nonlinearly generated wave and the incident one [33], which is at the basis of quasi phase matching technique. As previously discussed, if the crystallites were to be randomly inverted along the axis of propagation of the incident wave, this could also lead to random quasi-phase matching.

# 4. Fabrication and linear characterization of ZnS nanowaveguides.

Following the fabrication of ZnS thin films on insulator described in section 2, we proceeded to the fabrication of nanowaveguides using electron-beam lithography, patterning waveguides from 500 nm to 1μm width. ZnS waveguides are then defined using reactive ion etching (RIE) with a $CH_4/H_2/Ar$ gas mixture (Figure 4.a), optimized at low power (50 W) and 0.04 mbar pressure, with flow rates of 6.3 sccm ($CH_4$), 50 sccm ($H_2$), and 8.3 sccm (Ar). The etching rate, determined via interferometric measurements, reaches 7 nm/s, ensuring high-resolution pattern transfer essential for low-loss, high-performance photonic structures. The etching selectivity of the hard mask in this process limits the processing of ZnS waveguides to thicknesses of 500 nm and below. Figure 4 a,b) shows the cleaved facet and the sidewall of a fully processed ZnS waveguide. We used Fabry-Perot (FP) fringe measurements [34] to assess the propagation losses in the ZnS guides at telecom and visible wavelength ranges. Measurements were carried out using tunable Littman-Metcalf lasers in the visible and infrared wavelengths, whose polarization was controlled in order to study propagation losses in the $TE_{00}$ and $TM_{00}$ mode. Injection and detection are done in free space using the experimental set up presented previously and data acquisition was performed using silicon and Germanium photodiodes respectively for the visible and infrared wavelength scans. The contrast of the Fabry-Perot fringes on the spectra shown in Figure 4 c,e allowed us to estimate average propagation losses lower than 55 dB/cm in the whole wavelength range and even lower than 20 dB/cm in some spectral regions (Figure 4 d,f). Average losses in the visible and infrared range are 55 and 16 dB/cm for $TE_{00}$ and 45 dB/cm and 40 dB/cm for $TM_{00}$ respectively. The origin of the propagation losses can be attributed to sidewall and surface roughness (Figure S2). The measured propagation losses are competitive with the ones of nanowaveguides from other more mature platforms and are compatible for second order nonlinear experiments [5].

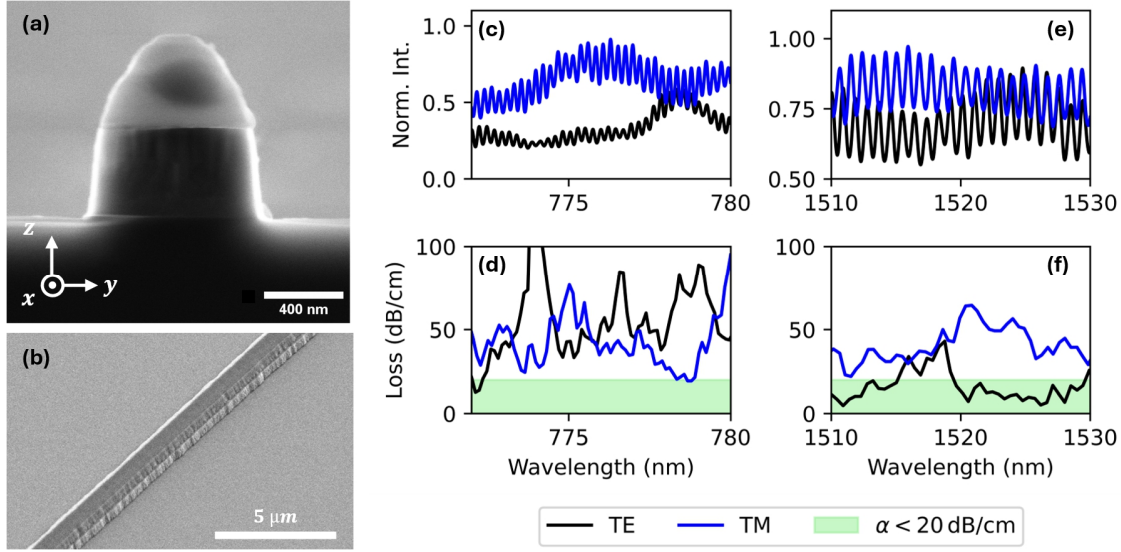

*Figure 4 SEM images of a) the facet and b) the sidewall of ZnS waveguide. c-f) Transmission spectra and propagation losses in visible and infrared region for TE and TM polarization. Values in green are less than 20 dB/cm.*

## 5. Second harmonic generation experiment

Second harmonic generation is performed in the optimal TM → TM polarization configuration, defined in the orthonormal reference frame shown in figure 4.a, involving the $d''_{33}$ element. For a waveguide 810 nm wide and 500 nm high, dispersion curves calculated for such an architecture with COMSOL show that the only possible modal phase matching from the $TM_{00}$ is reached at a wavelength of 1480 nm with the $TM_{02}$ mode (Figure 5.a). Under these conditions, the maximum theoretical conversion efficiency in CW regime, contingent on optical losses, waveguide length, and modal dispersion for a ZnS monocrystalline waveguide, reaches 21 %/W/cm² (see supporting information for more details). The experiment was conducted by injecting a picosecond pulse from an optical parametric oscillator (OPO) into the waveguide with an optical power of 8.2 mW and a repetition rate of 82 MHz. The OPO signal is transmitted to the laboratory via a 20 m optical fiber link. By operating at low optical power, nonlinear effects in the fiber are minimized, though the resulting chirp remains uncompensated. Polarization is precisely controlled at both injection and detection stages.

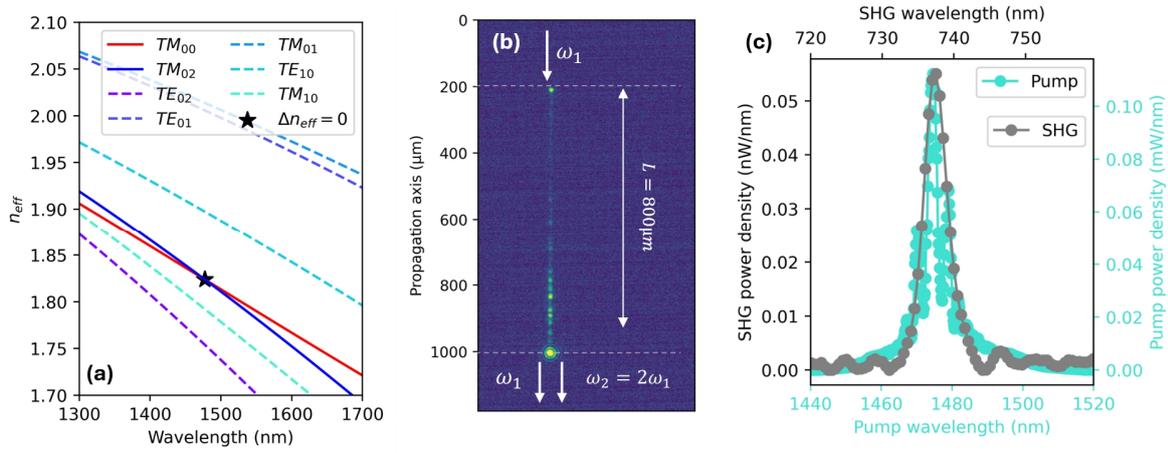

*Figure 5 a. Dispersion curve of several mode supported in the used waveguides geometry. b. Top view of SHG-generated signal scattering at 737 nm. The pump is injected from the top and the nonlinear signal grows during propagation. c. Pump and second harmonic spectrum (respectively cyan and gray plots).*

The OPO's wavelength was scanned from $\lambda = 1.1$ µm to $\lambda = 1.6$ µm to identify the optimal wavelength at which the SH signal is maximal, confirming a phase-matched process with a sharp peak at a wavelength $\lambda = 1474$ nm, only slightly different from the theoretically calculated one. A top-view microscope with a visible camera was used to image the scattered SHG light from the surface. As shown in Figure 5b, the microscopy image revealed the characteristic signal build-up, further validating phase-matched SHG. Spectra recorded with two optical spectrum analyzers in the telecom and visible ranges, as presented in Figure 5.c, confirm second harmonic generation at 737nm. These observations demonstrate that the TM →TM nonlinear process can be clearly observed in spite of the uncontrolled twist nonlinear process. The observation of a phase matching condition at the expected spectral position for monocrystals indicates i) the absence of random quasi-phase matching, which is expected due to the size of the crystallites [34] ii) a preferential orientation of the crystallites along one of the [111] crystal direction. Further efforts in ZnS crystal polarity engineering should enable demonstration of high-efficiency frequency conversion with ZnS monocrystalline devices or in the opposite broadband frequency conversion using random QPM in ZnS polycrystalline structures with tailored domain size. The instantaneous conversion efficiency in pulsed regime can be estimated to 0.4 %/W/cm² using the assumption of rectangular pulses (see supporting information for measurement and calculation detail). This evaluation should be close to the conversion efficiency in the CW regime. The discrepancy between theoretical and experimental values may be attributed to the orientation distribution of the crystallites, suggesting that 60% of the crystallites share a similar [111] crystal orientation [35]. This work represents an important milestone, as it reports the first observation of second harmonic generation (SHG) in ZnS nanowaveguides with a conversion efficiency close to that measured for the same modal phase matching in other photonic platforms such as GaP [36]. Future investigations will focus on nonlinear spectroscopic studies of this distribution in ZnS nanowaveguides, using a next-generation platform designed to enable phase matching within the spectral range of an available CW laser paving the way for further advancements, where optimizing waveguide length and reducing surface roughness will enhance nonlinear conversion efficiency.

## 6. Conclusion

In conclusion, we have realized the first ZnS nanowaveguides and demonstrated SHG in this platform with a first estimation of SHG conversion efficiency, marking a significant milestone in II-VI nonlinear integrated photonics. The nonlinear response of these polycrystalline waveguides has been theoretically analyzed and validated through SHG observations, in agreement with our calculations. Future efforts will focus on optimizing waveguide design to enhance nonlinear conversion efficiency, particularly for frequency conversion at shorter wavelength in the visible range and in the mid-infrared. Additionally, advanced polishing techniques will be explored to further reduce optical losses. By improving the fabrication and decreasing losses down to 5dB/cm, we forecast an increase in nonlinear conversion efficiency of two orders of magnitude with 1.2 cm long waveguides.

**Supporting Information**.

Supporting information provides experimental details and discussion about the propagation loss origin and the second harmonic generation experiment (PDF).

**Corresponding Author**

antoine.lemoine@insa-rennes.fr

**Author Contributions**

Author 1 was responsible for manuscript writing, device fabrication, ellipsometric measurements, simulations, and optical characterization. Author 4 performed the XRD measurements. Authors 5 and 6 contributed to the deposition of ZnS thin films. Author 7 assisted with tensor calculations. Authors 2, 3, 8, 9, and 10 provided help on manuscript preparation. Author 11 supervised the study.

**Acknowledgement**

The authors acknowledge nanoRennes for the technological support, a platform affiliated to RENATECH+ (the French national facilities network for micro-nanotechnology).

**Funding sources**

This research was supported by "France 2030" with the French National Research agency OFCOC project (ANR-22-PEEL-0005).